\documentclass[twocolumn,preprintnumbers,amssymb,amsmath,9pt]{revtex4}[10pt]
\usepackage{graphicx}% Include figure files 
\usepackage{dcolumn}% Align table columns on decimal point 
\usepackage{bm}% bold math 
\begin{document} 
\input epsf.tex
\newcommand{\beq}{\begin{eqnarray}}% can be used as {equation} or {eqnarray}
\newcommand{\eeq}{\end{eqnarray}}
\newcommand{\nn}{\nonumber}
\newcommand\T{\rule{0pt}{2.6ex}}
\newcommand\B{\rule[-1.2ex]{0pt}{0pt}}
\def\ltap{\ \raise.3ex\hbox{$<$\kern-.75em\lower1ex\hbox{$\sim$}}\ }
\def\gtap{\ \raise.3ex\hbox{$>$\kern-.75em\lower1ex\hbox{$\sim$}}\ }
\def\CO{{\cal O}}
\def\CL{{\cal L}}
\def\CM{{\cal M}}
\def\tr{{\rm\ Tr}}
\def\CO{{\cal O}}
\def\CL{{\cal L}}
\def\CM{{\cal M}}
\def\mpl{M_{\rm Pl}}
\newcommand{\bel}[1]{\be\label{#1}}
\def\al{\alpha}
\def\bt{\beta}
\def\eps{\epsilon}
\def\eg{{\it e.g.}}
\def\ie{{\it i.e.}}
\def\mn{{\mu\nu}}
\newcommand{\rep}[1]{{\bf #1}}
\def\be{\begin{equation}}
\def\ee{\end{equation}}
\def\bea{\begin{eqnarray}}
\def\eea{\end{eqnarray}}
\newcommand{\eref}[1]{(\ref{#1})}
\newcommand{\Eref}[1]{Eq.~(\ref{#1})}
\newcommand{\gsim}{ \mathop{}_{\textstyle \sim}^{\textstyle >} }
\newcommand{\lsim}{ \mathop{}_{\textstyle \sim}^{\textstyle <} }
\newcommand{\vev}[1]{ \left\langle {#1} \right\rangle }
\newcommand{\bra}[1]{ \langle {#1} | }
\newcommand{\ket}[1]{ | {#1} \rangle }
\newcommand{\ev}{{\rm eV}}
\newcommand{\kev}{{\rm keV}}
\newcommand{\Mev}{{\rm MeV}}
\newcommand{\gev}{{\rm GeV}}
\newcommand{\tev}{{\rm TeV}}
\newcommand{\mev}{{\rm MeV}}
\newcommand{\meV}{{\rm meV}}
\newcommand{\mnu}{\ensuremath{m_\nu}}
\newcommand{\nnu}{\ensuremath{n_\nu}}
\newcommand{\mlr}{\ensuremath{m_{lr}}}
\newcommand{\acc}{\ensuremath{{\cal A}}}
\newcommand{\mav}{MaVaNs}
\newcommand{\disc}[1]{{\bf #1}} 
\newcommand{\mh}{{m_h}}
\newcommand{\hb}{{\cal \bar H}}
\newcommand{\me}{\mbox{${\rm \not\! E}$}}
\newcommand{\met}{\mbox{${\rm \not\! E}_{\rm T}$}}

\title{Nonstandard Higgs Decays with Visible and Missing Energy}
\author{Spencer Chang}
\affiliation{Center for Cosmology and Particle Physics,
  Dept. of Physics, New York University,
New York, NY 10003}
\author{Neal Weiner}
\affiliation{Center for Cosmology and Particle Physics,
  Dept. of Physics, New York University,
New York, NY 10003}
\preprint{}
\date{\today}
\begin{abstract}
Experimental and theoretical clues both suggest that the Higgs boson has a mass below the LEP2 lower limit of 114.4 GeV.  If true, this suggests that the dominant Higgs decay is nonstandard while the production cross sections remain unmodified. We consider the possibility of nonstandard Higgs decays in the language of On-Shell Effective Theories (OSETs), and discuss a little considered class of Higgs decays, with a topology of both visible and missing energy.  We study existing LEP constraints on such decays, and find that such decays would in general be allowed experimentally for $\sim$ 100 GeV mass Higgses. Simple model realizations of these decays exist, which can occur in supersymmetric models and also in models with additional massive neutrinos. Some potential searches that can be performed at Tevatron and LHC, contained in standard supersymmetry topologies of leptons and missing energy, offer the possibility of discovering such Higgses.      

\end{abstract}

\maketitle

\section{Introduction}
Over the past thirty years, an impressive array of experiments have found agreement with the Standard Model (SM) of particle physics with included neutrino masses.  There have been no signals from direct searches and essentialy no significant deviations in precision measurements, a fact that is unexpected due to our theoretical priors.   The hierarchy problem, for example,  suggests that new physics is required to stabilize the electroweak scale.  In general this new physics generates deviations from the SM that should have already been discovered, which has lead to some uneasiness in terms of our theoretical understanding in electroweak physics.    

However, it is also true that the SM has yet to be fully confirmed, in that the Higgs boson remains undiscovered.  Within the SM, the mass of this particle is unknown, but once specified, all of its physical properties are predicted.  The electroweak precision observables (EWPO), accurately measured at LEP and other colliders, put an indirect upper bound on the Higgs mass of 144 GeV at 95\% CL with a central value of 76 GeV \cite{lepewwg}.  The newest measured top and $W$ mass have both contributed in lowering the upper bound.  This has started to be at odds with the direct search limit of 114.4 GeV at 95\% CL set by LEP2 \cite{Barate:2003sz}.

There has also been some tension in beyond the Standard Model (BSM) theories that contain Higgs bosons.  For example, the Minimal Supersymmetric Standard Model (MSSM) suffers from a little hierarchy problem, with an expected few percent level tuning required if the Higgs mass is above 114.4 GeV.  In light of these issues, people have readdressed what this conflict entails in the Higgs sector.  
There is the traditional approach that the Higgs is heavier than 114.4 GeV and the slight disagreement with the indirect limit is either due to chance or to new contributions to the EWPO \cite{Han:2004az,Grojean:2006nn}.  However, it is important to emphasize that the direct and indirect limits probe different physical properties of the Higgs; in fact, a Higgs with slightly nonstandard properties can be consistent with both. In particular, having a Higgs with the usual couplings to SM particles preserves the standard Higgs production cross section and is consistent with EWPO if the Higgs mass is light.  At the same time, direct search limits are avoided if there are new decays for the Higgs that dominate over the standard decays.  Another aspect of the Higgs which enables these scenarios is that for the Higgs masses probed at LEP, the Higgs decay width into Standard Model modes is quite small, since the bottom quark Yukawa is so weak.  Thus if there are new light states below the Higgs mass, the Higgs is prone to dominantly decay into these new modes.   The hints from theory and experiment and this particular susceptibility of the Higgs sector suggest that we should take such a nonstandard Higgs scenario seriously, which we will do for this paper.

In this regard, there has been great interest in Higgses that decay nonstandardly \cite{Dermisek:2005ar,Dermisek:2005gg,Chang:2005ht,Schuster:2005py,Graham:2006tr,Carpenter:2006hs,Chang:2006bw,Dermisek:2006wr,Dermisek:2006py,Graesser:2007pc,Graesser:2007yj,deGouvea:2007uz}.   In this note, we will extend the scope of considered decays by exploring nonstandard decays with a topology containing both visible and missing energy. Such decays can occur naturally in theories beyond the Standard Model, and it is important to consider what constraints exist within current data.  Moreover, since current and future Higgs searches, focusing on Standard Model Higgs decay products, are weakened by reduced statistics, it is also crucial to investigate how Higgs searches should adapt to discover these nonstandard decays. 

A natural language to describe the phenomenology is in terms of On-Shell Effective Theories (OSETs).  In the next section, this allows us to give a general discussion of the relevant phenomenology without couching it within a specific model realization.  In particular, we discuss what LEP2 results apply to this scenario and what parameter space is allowed.  In section \ref{sec:model}, we will give examples of realizations of this new Higgs phenomenology to demonstrate in what new physics models these could appear.   In section \ref{sec:searches}, we then give a preliminary discussion of potential searches for these Higgs decays that can be performed at Tevatron and LHC.  Finally, in section \ref{sec:conclusions}, we conclude.

\section{Evading the LEP Limits with NonStandard Higgs Decays}
Before we can address what modes allow for a Higgs lighter than the Standard Model limit from LEP, we must consider what studies have been performed. (See \cite{Chang:2005ht} for an extended discussion.) At LEP, the (SM) Higgs is produced via Higgsstrahlung  from a Z boson. The topology then studied involves both the decay of the Z as well as the Higgs. Easily studied $e^+e^-$ and $\mu^+\mu^-$ decays of the Z are infrequent, and thus most analyses use the two jet decays of the Z, or the invisible (neutrino) mode. 

For a Higgs produced with SM strength, strong limits exist for it decaying into two b-jets (115 GeV) \cite{Barate:2003sz}, two taus (115 GeV) \cite{Barate:2003sz}, two unflavored jets (113 GeV) \cite{:2001yb}, neutrinos or other invisible particles (114 GeV) \cite{:2001xz}, and two photons (117 GeV) \cite{Rosca:2002me} where the limits assume that the decay is 100\% into the modes in question. Decays into muons or electrons would also be at the kinematical limit. As a consequence, the full gamut of SM two body decays is essentially as strongly constrained as the SM Higgs limit. Thus, viable nonstandard Higgses must either decay into a multibody final state directly or must first decay into a two-body state that involves some BSM particle, which then ultimately decays into Standard Model states. Since phase space suppression would require strong couplings in the former case, we choose here to focus on the ``cascade decays'' of the latter.

LEP studies constrain states lighter than $\sqrt{s}/2$ to be neutral, essentially all particles into which the Higgs could decay. Thus, the BSM states introduced for the cascade decay must be neutral.  The simplest case would be for the particle to be stable, but this would then be constrained by invisible Higgs decays. Thus, the simplest unconstrained case must involve further decays. Once this is considered a multitude of possibilities open up, and it becomes important to determine how to discuss these questions in a suitably model independent way.

\subsection{OSETs and Appended OSETs}
Recently, a means of discussing certain experimental signatures without reference to a specific Lagrangian has been introduced via the language of On-Shell Effective Theories (OSETs) \cite{Arkani-Hamed:2007fw}.  In these descriptions, the phenomenology reduces to the physical parameters of production rates, branching ratios and masses, where offshell particles have been ``integrated out,'' allowing a discussion of signatures without reference to a specific model. In the present situation, this is ideal, because one prefers to understand the constrained signatures, in order to better understand what models might yield them. However, it is often difficult to choose an OSET in a vacuum, without any motivation. Here, we have a simple motivation: namely, to take the SM Higgs sector and ``append'' new states near the bottom of the spectrum.  Thus, in the nonstandard Higgs scenario, production rates are not modified, only the decay chains are altered.

\begin{figure}
\includegraphics[width=4cm]{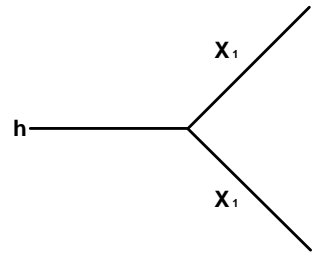}
\caption{Higgs decay into a pair of light neutral particles. \label{fig:oset1}}
\end{figure}

The simplest OSET which evades the LEP limit is shown in figure \ref{fig:oset1}. Here, the Higgs decays into two particles, both of which presumably undergo two-body decays into SM states. The case where the intermediate particle is a pseudoscalar, decaying into two b-jets was considered as a means of providing a natural model in \cite{Dermisek:2005gg}, but because the two-b and four-b final state analyses are not independent, such scenarios are, in fact, nearly as constrained as the SM (see the discussion in \cite{Chang:2005ht}). Consequently, decays where the SM final state is four taus have been considered \cite{Chang:2005ht,Dermisek:2005gg,Graham:2006tr,Schuster:2005py}, as the LEP limits on the Higgs mass do not extend above 86 \gev. Such a scenario is still significantly tuned \cite{Schuster:2005py}, but not as badly as the MSSM \cite{Dermisek:2006wr}. Other possibilities exist within this OSET, such as four light jets \cite{Chang:2005ht,Chang:2006bw}, or six light jets \cite{Carpenter:2006hs}, which may be studied at the LHC \cite{Chang:2006bw,Martin:2007dx,Kaplan:2007ap}.  However, because of the strong constraints on these OSETs, it is worthwhile to consider the next simplest possibility as well. As we shall discuss, such possibilities arise naturally in theories beyond the Standard Model.

The next simplest possibility involves two states lighter than the Higgs, with one unstable and one stable. The lightest stable neutral particle ($X_1$) could be the dark matter, or it could be the Standard Model neutrino. The heavier ($X_2$) could be a neutralino, sneutrino, sterile neutrino, or any other new neutral particle. With two states lighter than the Higgs, this opens the possibility that the Higgs could decay off diagonally, i.e., $h\rightarrow X_2 X_1$, where $X_2$ subsequently decays. Such a possibility is intriguing because a Higgs decay involving this state would contain both missing and visible energy. Because nearly all Higgs studies at LEP require momentum conservations (notable exceptions being the invisible Higgs search \cite{:2001xz}, $h\to WW^*$ search \cite{Abdallah:2003xf,Schael:2006ra}, and the OPAL decay independent analysis \cite{Abbiendi:2002qp}), such a decay would not be picked up by the strongest Higgs searches, and generically only the relatively weak decay independent limit of 82 \gev\ applies \cite{Abbiendi:2002qp}.

\subsection{Phenomenology of Higgs Decays with Missing Energy \label{sec:pheno}}

\begin{figure}
\includegraphics[width=4cm]{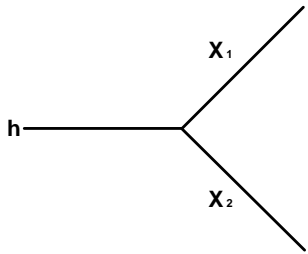}
\caption{Higgs decay into a heavy-light pair of neutral particles. \label{fig:hdecay}}
\end{figure}

In order to consider the possible constraints, we must specify the decays under study. We will refer to the two light neutral states as $X_{1,2}$ where $M_2>M_1$. It is possible that the decay would proceed via $h\rightarrow X_2 X_2 \rightarrow X_1 X_1 SM$, but there is often little kinematical phase space for such decays. Our focus will be on the decay $h\rightarrow X_2 X_1 \rightarrow X_1 X_1 SM$, see Fig. \ref{fig:hdecay} \footnote{Some recent papers have looked at a subset of such decays, although they typically do not have these decays dominate for Higgses below the LEP2 limit.  Of the type $h \to 2 X_2$, Graesser has motivated massive right handed neutrinos at the weak scale which allows such decays.  See \cite{Graesser:2007pc, Graesser:2007yj} for more about the model and limits and \cite{Aranda:2007dq} for related work.  For the off diagonal decay $h \to X_2 X_1$, De Gouvea has recently looked at such decays with heavy neutrinos, but finds that they cannot dominate due to neutrino constraints \cite{deGouvea:2007uz}.  We will give a model in one of the later sections that does allow such decays to dominate.}. In terms of the model realizations that we will discuss later, the particles are listed in table \ref{table:oset}.  It is our assumption that these new decays dominate over the Standard Model decay modes.  In fact, the SM Higgs search requires that the nonstandard decay be at least 75\% of the branching ratio for a Higgs of 100 GeV.  This reduction of SM decays reduces the reach of standard searches and thus, in order to maintain discovery at the LHC for such Higgses, it will be important to study this nonstandard phenomenology in more detail.    

Let us begin by repeating our assumptions: the common characteristics are that $X_{1,2}$ are lighter than the Higgs where $M_1 + M_2 < M_h$, are electrically neutral and that $X_2$ is unstable.  On the other hand, their spin, gauge representations and particle names vary amongst the possibilities.  Since the Higgs boson is a scalar, at this point, the decay kinematics are only set by the masses of $X_1, X_2$ and do not depend on the spin of the particles.      

\begin{figure}
($i$) \includegraphics[width=4cm]{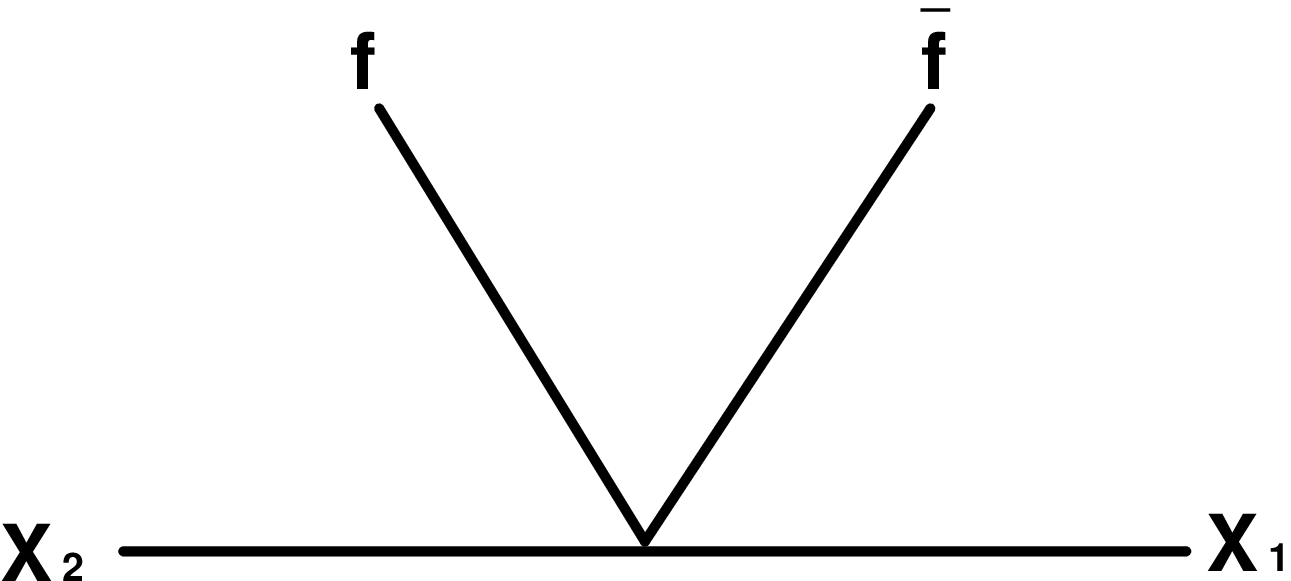}\\[.4in] ($ii$) \includegraphics[width=4cm]{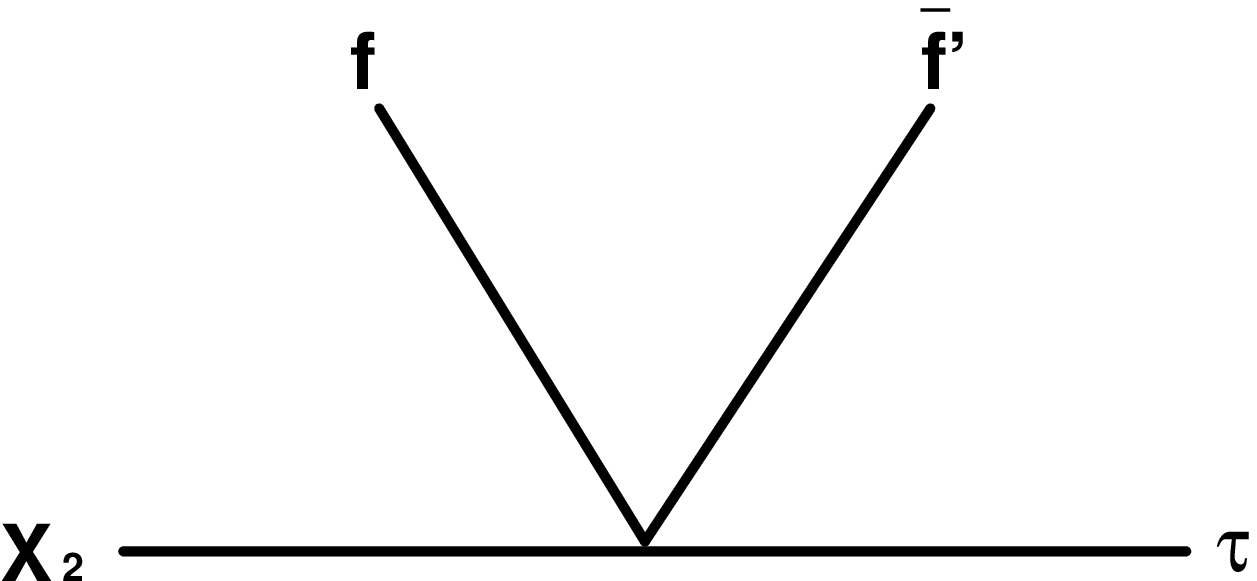} 
\caption{$X_2$ three body decays into two SM fermions and $X_1$ or $\tau$. These are often mediated by offshell gauge bosons, $i$) $Z^*$ and $ii$) $W^*$ respectively.  \label{fig:x2gauge}}
\end{figure}

\begin{figure}
($iii$) \includegraphics[width=4cm]{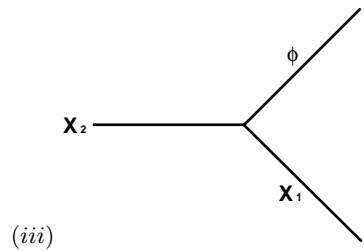}
\caption{$iii$) $X_2$ decay into $X_1$ and an onshell scalar $\phi$.  Depending on it's mass, the scalar usually then decays into $b\bar{b}$ or $\tau\bar{\tau}$.  \label{fig:x2scalar}}
\end{figure}

The complete decay topology is then determined by the decay of $X_2$.  These decays are listed in table \ref{table:oset} and depicted in Fig. \ref{fig:x2gauge} and \ref{fig:x2scalar} with the labels $i, ii, iii$.  Most $X_2$ decays we consider are into $X_1+SM$ (type $i, iii$), while a special case is into $\tau f \bar{f}'$ (type $ii$ which can arise via an offshell $W^*$).

In general, many of these decays will occur simultaneously, i.e., if the decays are mediated by off-shell gauge bosons.
% The offshell $Z^*, W^*$ are only shorthand for the fact that the $X_2$ decay down to $X_1$ (or $\nu_\tau$ in the neutrino case) is a three body decay mediated by the offshell gauge boson. %
In terms of an actual OSET, it's unimportant to discuss the off-shell mediator of these decays, but we shall often reference them as motivated sets of branching ratios that are particularly interesting to consider. Given that there are potentially other mediators of these three body decays (such as sleptons, squarks, etc.), it is important to keep in mind that the branching ratios are not always set by those of $W^*,Z^*$.     
% \disc{TMI} In the neutralino case, there is also the possibility of an onshell scalar $\phi$ which can be a scalar or pseudoscalar, which will predominantly decay into $b$ quarks if heavier than two $B$ mesons and otherwise go in two $\tau$'s.

The decay phenomenology one might expect from specific models is simply summarized.   For the decays mediated by an offshell gauge boson, the phenomenology is predominantly into two light quark jets and at a subdominant rate into leptons.  For the onshell scalar decay, there is predominantly 2$b$ or 2$\tau$ decay, with a mass peak at $m_\phi$.  In each decay, missing energy is carried off by $X_1$'s or neutrinos and visible energy is of the form of SM fermions.  With such simple phenomenologies, we can deduce the LEP2 constraints on such decays.         

\begingroup
\squeezetable
\begin{table}
\begin{tabular}{|l|l|l|l|l|}
\hline
\T Model & $X_1$ & $X_2$ & $X_2$ Decay & Class \\[.1cm]
\hline 
\T Neutrinos & $\nu_\tau$ & $\nu_{\text H}$& $\nu_{\text H} \to Z^* + \nu_\tau, W^* + \tau $& $i, ii$\\[.1cm]
Neutralinos & $\chi_1$ & $\chi_2$ & $\chi_2 \to \phi + \chi_1, f\bar{f} + \chi_1 $& $i, iii $\\ [.1cm]
Sneutrinos & $\tilde{\nu}_1$ & $\tilde{\nu}_2 $& $\tilde{\nu}_2 \to f\bar{f} + \tilde{\nu}_1$ & $i$ \\ 
\hline 
\end{tabular}
\caption{OSET Particles, Decays and Potential Model Realizations \label{table:oset}}
\end{table}
\endgroup

\subsection{Current Higgs Search Limits}
Because of the number of possibilities, it is important to be systematic in our approach to the limits on such senarios. 
As already stated, the Higgs decay topology is a combination of both visible decay products from Standard Model fermions and an invisible component from $X_1$'s and/or neutrinos.  Thus, we shall limit ourselves to Higgs decays $h \to f \bar{f} + \me$ and $h \to f \bar{f}' l + \me$, which have not been directly searched for by LEP collaborations in a systematic fashion.  Because many of the searches put their limits in terms of production cross section, we shall first detail the relevant studies and their limits.  After this, we shall relate those cross section limits to Higgs production cross sections, which can be non-trivial.  Due to this issue, we will discuss dedicated Higgs searches first, as this conversion is straightforward.  

However, before we discuss in detail the LEP2 constraints, it is important to give some caveats on such an analysis.  First of all, we are largely assuming that one can apply cross section bounds for a different signal to our Higgs decay.  That is, we will consider the topology only, without consideration of the specific kinematic features. This assumes that the efficiencies for both types of signals are similar.   More importantly this also assumes that the analyses are merely counting experiments (in a single bin) with no information about kinematical discriminating variables being used in the analysis.  These will make our constraints a bit uncertain and perhaps conservative, but hopefully still applicable to a variety of models.  Finally, to deduce the constraints a non-Higgs analysis has on the Higgs decay, we must completely specify the Higgs decay topology.  This involves both the onshell Z decay as well as the specific $X_2$ decay.

%\begingroup
%\squeezetable
\begin{table*}
\begin{tabular}{|c|c|c||c|l|}
\hline
\T $X_2 {\rm Decay} $ & Type & Higgs Decay Topology & $Z$ Decays Used in Limit & Limits\\[.1cm]
\hline 
\hline
\T anything & $i,ii,iii$ & anything + $\me$ & $l \bar{l}$ & OPAL Decay Independent [$m_h >$ 82 GeV] \\[.1cm] 
\hline
\T&&& &  LEP2 Invisible Higgs search \\[-.1cm]
 $\nu \bar{\nu} + X_1$ & $i$ & $\me$ & $l\bar{l}, q\bar{q}$ &   \\[-.2cm]
\T&&&& [BR $< (.17,.24,.41)$ for $m_h = (100,105,110)$ GeV]\\[.1cm]
\hline
\hline
\T $l \bar{l} + X_1 $ & $i,ii,iii $ & $l \bar{l} + \me $ & $l\bar{l},\nu \bar{\nu},q \bar{q}$ & ALEPH $WW^*$ [.044 pb] \\ [.1cm]
\hline
\T $l q \bar{q}'$ & $ii$ & $l q \bar{q}' + \me $ & $l\bar{l},\nu \bar{\nu},q \bar{q}$ & ALEPH $WW^*$ [soft jets .041 pb, hard jets .066 pb] \\[.1cm]
\hline
\T $\tau \bar{l} + X_1$ & $i,ii,iii$ & $\tau \bar{l} + \me $& N/A & No meaningful limit \\[.1cm]
\hline
\T $\tau q \bar{q}'$ & $ii$ & $\tau q \bar{q}' +\me$& $l\bar{l},\nu \bar{\nu},q \bar{q}$ & ALEPH $WW^*$ [.47 pb] \\[.1cm]
\hline
\T $\tau \bar{\tau} + X_1$ & $i,ii,iii$ & $\tau \bar{\tau} + \me$& $\nu \bar{\nu}$ & LEP2 stau search [.05 pb] \\[.1cm]
\hline
\hline 
\T \B $q \bar{q} + X_1$ & $i,iii$ & $q\bar{q} +\me$ & $\nu \bar{\nu}$ & LEP2 Squark Search [Bottoms .02 pb, Light Quarks .06 pb] \\
\hline
\end{tabular}
\caption{Higgs decay topologies for $e^+e^- \rightarrow hZ$, $h\to X_2 X_1$ as determined by decay of $X_2$.  Last two columns denote the limits placed by LEP2 searches with search channels set by the onshell $Z$ decay.  (Note: $l$ stands for electron or muon)  \label{table:higgsconstraints}}
\end{table*}
%\endgroup

\subsubsection{Higgs Searches Constraining Missing Energy Higgs Decays}
There were a few dedicated Higgs searches at LEP2 that were sensitive to Higgs decays containing missing energy.  
First of all, the decay independent Higgs search by OPAL applies, since it only focused on reconstucting the charged leptonic decay of the $Z$ accompanying the Higgs \cite{Abbiendi:2002qp}.  This gives a lower bound on the Higgs mass of 82 GeV, when the Higgs is produced at the Standard Model rate.  In the case where the ultimate decay is completely into stable neutral particles, the LEP2 search for invisible Higgs decay applies, which also is strongly constrained.  This would constrain the invisible decays of $X_2$.  As mentioned earlier, a Higgs that decays exclusively into invisible states is constrained to be heavier than 114 GeV.  However, if this occurs with a branching ratio of 20\%, as in the specific case of a mediating $Z^*$, a 104 GeV Higgs would still be allowed \cite{:2001xz}.  These limits are tabulated in Table \ref{table:higgsconstraints}.     

In principle, LEP2 searches for topologies of a visible Higgs decay with the invisible decay of the associated $Z$ could constrain topologies where instead the Higgs decays with missing energy.  However, the amount of missing energy is generally much less in the nonstandard decay compared to that coming from the $Z$ decay, thus we expect those searches to have reduced efficiency.  Another dedicated Higgs analysis which allows missing energy in the Higgs decay is the fermiophobic Higgs search $h \to WW^*$ within the leptonic channels.  These analyses \cite{Abdallah:2003xf,Schael:2006ra} will give us some of the strongest constraints on decays with leptons, so we will now address how we interpret their constraints on the nonstandard Higgs decay.

$WW^*$ decays require leptons in order to have true missing energy from neutrinos.  We will assume for our purposes, that those leptons and taus are not lost down the beampipe or misidentified as jets; making this simplification, these analyses will only constrain  nonstandard Higgs decays containing charged leptons.  We choose to adopt the ALEPH analysis \cite{Schael:2006ra} over the DELPHI analysis \cite{Abdallah:2003xf} for the following reasons.  First of all, the DELPHI analysis is over a smaller set of integrated luminosity and thus would be expected to have less of a reach.  The DELPHI analysis also employs neural net techniques to enhance signal over background; these techniques are highly signal dependent and thus are suspect for the model independent applications we are envisioning.  Plus, ALEPH provides enough information to reconstruct their efficiencies for $WW^*$ decay products  which, since they only employ some simple cuts, should more reliably serve as a proxy for the efficiencies of the nonstandard decay.  Unfortunately, there is a tradeoff since the ALEPH analysis constrains fewer decay modes, but in common search channels, DELPHI constraints are similar to those obtained from ALEPH, so there isn't a significant degradation in the constraints.      

There are some standard Higgs searches which would seem to strongly constrain the nonstandard decays.  For instance, the $h\to 2\tau$ search might seem to constrain the type $i$ and $iii$ decays, where the fermions produced are taus.  We will focus on the scenario of the onshell scalar decay $iii$, where the scalar is lighter than the $b$ threshhold, as this is the scenario where the tau decays are expected to be $O(1)$.  There are a few distinctive kinematic features of this nonstandard decay which should make the standard search ineffective.  First of all, the onshell scalar is often boosted since we expect $m_{\phi} \sim 4-8 \; \gev$, giving a typical boost $\gamma \sim m_{X_2}/(2 m_{\phi}) \sim 5$ where we have taken the mass of $X_2$ to be 60 GeV.  This means that the taus are separated by less than 30 degrees about 65\% of the time.  This will cause the tau identification to fail, since the taus are often in each other's isolation cone.  Some analyses also use selection cuts that require the taus to be separated by as much as 70-90 degrees.  This does not cut most of the standard Higgs decay since there the taus are often back to back, but is a very strong cut on the nonstandard decay.  This distinctive difference in kinematics should also cause problems for the kinematic fitting constraints and likelihoods that are used in the analyses.  For these reasons, it does not appear that the standard $h \to 2\tau$ search puts the strongest constraints on this decay;  in the next subsection, we will instead use the stau search limits, which are arguably more reliably applied to our scenario. 

\begin{figure}
\begin{center}
\includegraphics[width=9cm]{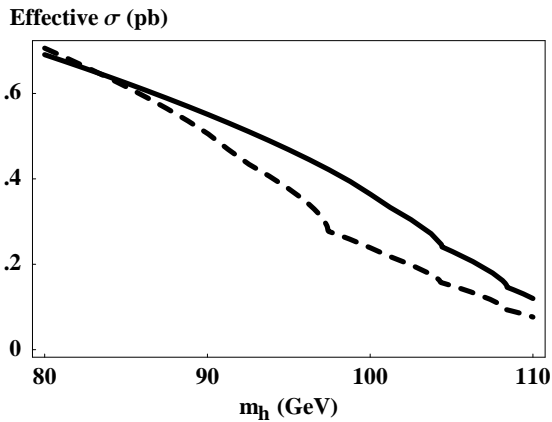}
\caption{Effective Higgs production cross section for squark and $WW^*$ limits (bold line) and stau limits (dashed line).  Note: to apply the squark and stau limits, one must also multiply this cross section by the branching ratio of the $Z$ invisible decay.  \label{fig:xseceff}}
\end{center}
\end{figure}

To deduce the cross section constraints on particular nonstandard Higgs decays, we use the simulated backgrounds and number of detected events within the different ALEPH search channels to deduce the constraints on the following Higgs decays: $h\to \tau q \bar{q}' +\me, l \bar{l} + \me, l q \bar{q} +\me, q \bar{q} l + \me$.  What distinguishes the last two entries is that the jets are softer and harder, respectively, since they come from the $W^*$ and $W$ respectively in the ALEPH search.  We mimic the LEP2 standard of $CL_s \equiv CL_{s+b}/CL_b \geq .05$ as the 95\% C.L. limits, but assume that all events are in a single bin which gives conservative (less stringent) limits.  We combine different channels that have the same Higgs decay with a different $Z$ decay and require $CL_s \equiv \Pi_i CL^i_s \geq .05$.  We assume the ALEPH efficiencies approximate the efficiencies for the nonstandard Higgs, and use this to generate a cross section limit on each decay; we then list these limits in Table \ref{table:higgsconstraints}.  In Fig. \ref{fig:xseceff}, we list the effective Higgs production cross section in bold.  All that remains is to multiply by the branching ratios of the Higgs and $X_2$ decay in question and compare to the cross secton limit in Table \ref{table:higgsconstraints}.        

\subsubsection{Non-Higgs Searches Constraining Missing Energy Higgs Decays}
In order to put constraints on the remaining decays with missing energy, we will have to turn to non-Higgs searches and in particular, searches for pair production of supersymmetric particles.  
%In Table \ref{table:higgsconstraints}, we have listed the relevant decay channels when $X_2$ decays into $X_1 f {\bar f}$ or onshell scalar $\phi$, which we have dubbed as neutral Higgs decay (decays $i$ and $iii$ respectively).  To the best of our knowledge, the strongest constraints come from the listed cross section limits from LEP2 searches for squark pair production.  
We will apply constraints by using the combined LEP2 cross section limits for sbottom, stop and stau searches \cite{lepcombined}.  The search that applies to each nonstandard decay is straightforward except where we utilize the stop search $\tilde{t}\to c \chi_1$ to constrain decays with light quarks.  The cross section limits for a given topology vary under the assumption of kinematical variables being considered (e.g., the squark mass), and we will take the ``bulk'' region to give  the relevant limits.  Alternatively, we could take the strongest limits throughout the parameter space, which typically limits the allowed cross section limits to be a factor of two smaller, but the ultimate conclusions of the viability of a light Higgs with such decays is not strongly sensitive to which limits we take.  Again, to connect cross section limits for a different search to the nonstandard Higgs, we have assumed that efficiencies for both are similar.  This was verified for the sbottom search and we assume it to be a good guideline in general.  These limits appear in Table \ref{table:higgsconstraints}.    

Having described limits on topologies similar to those from the nonstandard Higgs decays, we must now make a connection to the production cross section of the Higgs. There are essentially two important elements.  First, we must take into account the branching ratio of the $Z$ from which the Higgs is radiated. I.e., it must decay invisibly in order to achieve the topologies in question from squark and slepton searches, hence there is an overall suppression of 20\% of the effective cross section \footnote{The case of the $X_2$ decaying invisibly is at least constrained by the invisible Higgs decay search.  In the particular case of the OPAL sbottom search, we have determined that the topology where the onshell $Z$ decays into bottom quarks and $X_2$ decays invisibly has neglibly small efficiencies compared to the onshell $Z$ invisible decay $X_2$ decay into bottom quarks.  This is because of the strong background of $WW, ZZ$ which in general makes this $X_2$ invisible decay topology inefficient under these searches.}. 

The second element is that
%Going back to non-Higgs analyses, another aspect of applying an analysis based on a different signal hypothesis is that it is necessary to ``convert'' cross section limits.  The strongest constraints we have on this Higgs scenario are searches for sparticle pair production at LEP2.
the sparticle searches are kinematically closed only when $2m_{sparticle} > \sqrt{s}$.  On the other hand, the Higgs production is closed when $m_h + m_Z > \sqrt{s}$, so the Higgs production can be closed when the sparticle production is still open.  In the sparticle searches we apply, the sparticle masses constrained allow production for all of the applied LEP2 accumulated luminosity.  This is not the case for Higgs production of masses above 90 GeV, so there is a reduced effective cross section for the Higgs scenarios.  The amount of reduction depends on the accumulated luminosity and the Higgs mass in question.  This leads us to consider the effective Higgs production cross section, which is essentially a luminosity weighted cross section.
\be
\sigma_{eff} = 1/{\cal L}_{\rm tot} \sum_i \sigma(s_i) {\cal L}_i
\ee
where $s_i$, ${\cal L}_i$ are the center of mass energy and luminosity for a given LEP2 run $i$. 
%The LEP2 integrated luminosities of $\sim 600$ pb$^{-1}$ ($400$ pb$^{-1}$) were acquired from $\sqrt{s}=182-208 \,\gev$ ($\sqrt{s}=192-208 \,\gev$), 
Taking this into account, we have plotted in Fig. \ref{fig:xseceff}, the effective Higgs cross section for squark and stau searches (bold and dashed lines respectively).  Note that these cross sections do not include the $Z$ invisible branching ratio of 20\% as detailed before, so one must first multiply by it to compare to the limits in Table \ref{table:higgsconstraints}.

 For example, to consider the limits on a 100 \gev\ Higgs decaying $h\rightarrow X_2 X_1 \rightarrow X_1 X_1 q {\bar q}$, the constraint is from squark searches, with a topology of two jets and missing energy. The effective Higgs cross section for this analysis is given by the solid line of Fig. \ref{fig:xseceff} which is .4 pb. Because the Z must decay invisibly to achieve this topology, the final effective cross section for this topology from Higgs production is roughly .08 pb, which is above the .06 pb limit from the  ``bulk'' region of the LEP2 squark search constraints.  Therefore, either the Higgs mass has to be raised or an additional branching ratio suppression has to exist for this nonstandard Higgs to be consistent with constraints.  
  
 \subsection{Summary of LEP Limits} 
% \disc{What about ditau search, does it apply to lnulnu or 2tauhad, tauhad lnu?}
 
Having reviewed the LEP limits and their relationship to Higgs production, it is worthwhile to stop and note what sorts of scenarios are presently allowed. The following is under the assumption that we apply the ``moderate'' (bulk region) sparticle limits and the ALEPH $WW^*$  limits listed in Table \ref{table:higgsconstraints}.
If the Higgs is produced with Standard Model strength, and decays via $h\rightarrow X_2 X_1 \rightarrow X_1 X_1 SM$, such a Higgs is allowed above $\sim 103\ \gev$ if 
SM = two light jets, $\sim 111\ \gev$ if SM = $b \bar b$,  $\sim 99\ \gev $ if 
SM = $\tau \bar \tau$, and $\sim 115\ \gev$ if SM = $e \bar e, \mu \bar{\mu}$.  If instead $X_2$ decays are visible, the Higgs has to be above $\sim 115\ \gev$ if $X_2\to l q \bar{q}'$ and $\sim 95\ \gev$ if $X_2 \to \tau q \bar{q}'$.  

Note these results are if the Higgs decays exclusively into one SM channel, which depends on the model realizations. 
%If, for instance, the decay went equally into $e \bar e$, $\mu \bar \mu$ and $\tau \bar \tau$, the limit would be $\sim 98\ \gev$. 
For instance, if the decay is mediated by an offshell Z, these limits are no longer constraining.  Instead the strongest constraint comes from the invisible Higgs search, which allows the Higgs to be heavier than 104 GeV if BR($h\to X_2X_1$)=1 and 98 GeV if BR($h\to X_2X_1$)=.8 (which is consistent with the SM Higgs search limit). 

Since $b$ quark limits are so strong, if the $X_2 \rightarrow X_1 SM$ decay proceeds via an onshell scalar, the situation is highly constrained if $\phi$ goes into $b$ quarks.  This requires that the Higgs be heavier than $\sim$ 111 GeV for moderate limits.  However, if there is an additional nonstandard decay for the Higgs, besides $X_2 X_1$, which reduces this branching ratio below 1/4, this is still allowed for a 100 GeV Higgs.   
To compare, in the case where $\phi$ is kinematically forced to go into two taus, the constraint only requires a Higgs mass of 99 %(107)%  
GeV for moderate %(strong)%
 limits with $h\to X_2 X_1$ order one.  

Another model realization, discussed in the next section, is where $X_{2,1}$ are (sterile and mostly tau respectively) neutrinos.  In this case, the component of the decay that is most strongly constrained is where $h\to l q \bar{q}' +\me$, with the constraint that the Higgs is heavier than 104 GeV for the soft jets limit and 100 GeV for the hard jets limit.  The kinematics of this nonstandard decay is somewhat in between both, so the true limit would probably be somewhere between.  

So to summarize the limits on certain model realizations, we have found that the offshell $Z$ scenario was most strongly constrained by the invisible Higgs search and required the Higgs to be heavier than 104 GeV.  In a model where $X_2$ decays into the onshell scalar $\phi$, if $\phi$ decays into taus the limit is 99 GeV whereas if it decays into $b$ quarks the limit is 111 GeV or requires a branching ratio of 25\% for a Higgs of 100 GeV.  Finally, for the scenario with a sterile neutrino, the decays mediated by offshell $W^*,Z^*$ had limits from the ALEPH $WW^*$ search requiring the Higgs to be heavier than 100-104 GeV depending on the jet kinematics of the $l q\bar{q}' +\me$ decay.

%If we instead consider the decay chain $h\rightarrow X_2 X_1 \rightarrow X_1 \tau SM$, (i.e., decay $ii$),  there is a rough limit of 100 GeV on the Higgs mass.  This suggests that these nonstandard Higgs decays can occur for a Higgs as light as 100 GeV with almost 100\% branching ratio if the decays are through offshell gauge bosons or into an onshell scalar that goes into $\tau$'s.  However, for onshell scalars that decay into $b$ quarks,  limits from the sbottom search require a reduced branching ratio $BR(h\to X_2 X_1) \lesssim 1/4 $ to allow a 100 GeV Higgs.         

Again, we should reiterate the caveats to these limits. In the analyses we have used to constrain the scenario, there are important steps taken due to certain signal properties which may not apply to the nonstandard Higgs.  For instance, the use of optimized cuts and likelihood distributions used for the original signals.  Not all analyses give enough information to recreate the analysis, so it is difficult to determine efficiencies these analysis steps have for the Higgs decay.  Indeed, we have assumed that such efficiencies are the same.  We have disregarded kinematical features by treating these searches as one bin counting experiments, making the limits conservative but hopefully model independent.  These issues suggest that our ultimate cross section limits are merely estimates and may be only correct at best up to factors of two.  However, absent a dedicated analysis for this Higgs decay, this is the best available estimate possible. 
%Moreover, it is unlikely that an analysis for the same topology as the Higgs decay, but optimized for different kinematics, would provide stronger constraints on the Higgs production, but this is a possibility.
 
\section{A Model Realization \label{sec:model}}
This phenomenology can arise simply in supersymmetric theories, for instance via $h\rightarrow \chi_2 \chi_1$ or $h\rightarrow \tilde \nu_2 \tilde \nu_1$ as listed in Table \ref{table:oset}. However, there are many associated issues, so we will defer those discussions to future papers. Here we present a simple model, which could arise in a variety of BSM situations, such as little Higgs theories or supersymmetry, in which the Higgs decays into a pair of neutrinos, i.e., $h\rightarrow \nu_2 \nu_1$, where $\nu_2$ is heavy and $\nu_1$ is approximately massless.

At first glance, one would think that a Higgs decay such as this would be impossible without significant tuning, because the large Yukawa coupling needed to compete with the standard decays. However, it is quite simple to construct a model in which there is a large Yukawa, but there remains a massless neutrino in the theory.

Consider the following interactions
\bea
{\cal L} = y_{\nu} H L N_R + \mu N_L N_R + h.c.
\eea
When the Higgs acquires a vev, there is a Dirac mass which marries the neutrino in $L$ to the neutrino in $N_R$. However, $N_R$ is already married to the neutral state in $N_L$. As a consequence, there are two left handed states, and only one right handed state. As all the masses are Dirac, one left handed state remains massless, which we identify with the SM neutrino.

However, this light state is not a purely Standard Model neutrino, which modifies couplings of the associated charged lepton to the $W$ boson, which are constrained from precision measurements. The strongest constraints come from the extraction of $G_f$ from $\tau \rightarrow e$ decays compared with $G_f$ from $\mu \rightarrow e$ decays as well as some EWPO, which constrain the mixing angles to be smaller than $O(10\%)$ \cite{Loinaz:2004qc, PDBook}.
Such constraints are the principal barrier to this new decay channel dominating; for e.g., with a 100 GeV Higgs the new decay can never be more than 50\% of the Standard Model decay width.  Plus it is also worth mentioning that the electroweak fit typically requires the Higgs to be heavier than the LEP2 limit \cite{Loinaz:2004qc}.  The necessary contributions to the EWPO can come from another sector of the theory, so we will choose to overlook this issue.  

\begin{figure}
\begin{center}
\includegraphics[width=9cm]{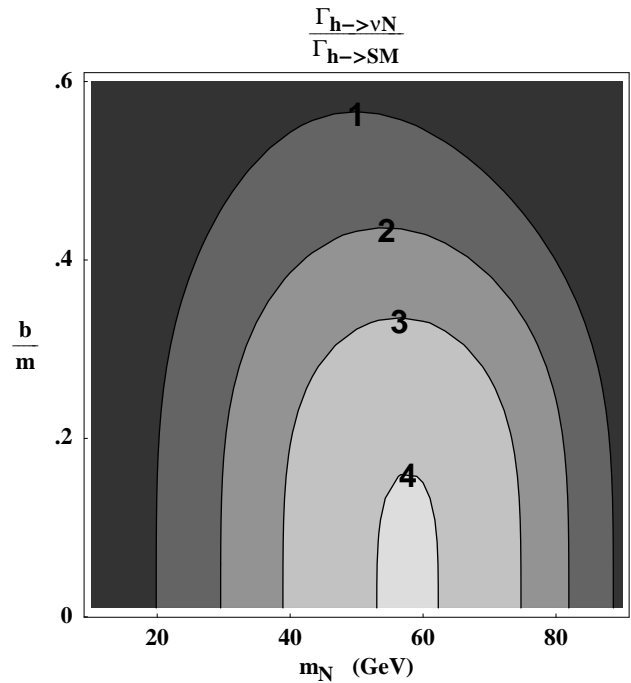}
\caption{Ratio of new decay width into neutrinos to that into SM decays $\Gamma_{\nu N}/\Gamma_{SM}$versus $b$ and mass of the heavy neutrino $m_N$, with values $\lambda = .082, m=120 \;\gev$.  Starting from the top (darkest) going down (lightest) the ratio is $< 1, 2, 3, 4$ respectively. \label{fig:Gneutrino}}
\end{center}
\end{figure}

A model which can allow the Higgs decay to dominate, consistent with constraints, is the following two Higgs doublet model
\bea
\nonumber {\cal L} &=& y_{\nu} \Phi L N_R + \mu N_L N_R + h.c. +\CL_{Yukawa}(H)\\[-.1cm]
&& \\[-.1cm]
\nonumber &-&  \lambda (|H|^2- v^2/2)^2 - m^2 |\Phi|^2 + (b^2 H^\dag \Phi  + h.c.) 
\eea
In the limit that $b^2 \lesssim v^2, \mu^2$, there is very little neutrino mixing with the sterile state, however there is a large Yukawa between the heavy light neutrinos due to a $\tan \beta$ like enhancement.  In particular this is possible now that there can be different mixing angles for the Higgs vevs and mass eigenstates.  As our requirements, we force the massless neutrino to have at most $\sin^2 \theta_\nu = 1\%$ sterile neutrino content and for $\vev{H}^2 +\vev{\Phi}^2 = v^2/2 \sim (175 \;\gev)^2$.  In fig. \ref{fig:Gneutrino}, we have chosen a value of $\lambda$ to get a light Higgs of mass 100 GeV and $m= 120 \; \gev$ and maximized the allowed neutrino mixing, plotting the contours of the ratio of nonstandard decay width to the standard one.  The contours go from top to bottom as $<1,2,3,4$ and it must be above 3 to be consistent with the Standard Model Higgs search.  This is highly dependent on the value of $m$ as a value above 125 GeV does not have any viable parammeter space while $m$ less than 115 GeV will start to have constraints from the SM search for the heavier Higgs boson.  The contour of 1 is also interesting since if there were three new massive neutrinos, this line could be consistent if all three neutrinos could attain maximum mixing.  However, the mixings would then be subject to separate constraints and in general cannot be as large as for $\nu_\tau$ \cite{Loinaz:2004qc}. 

\section{Future Higgs Searches at Hadron Colliders \label{sec:searches}}
The shutdown of LEP2 in preparation for LHC's start up has transferred the mantle of finding the Higgs to the Tevatron and LHC.  At this point, reanalyzing LEP2 data is not a realistic possibility for discovery.  Since these nonstandard decays are by definition inefficiently picked up by Higgs searches at LEP2, the crucial weakness of the collider was insufficient numbers of produced Higgses.  Thus, at best a LEP2 reanalysis would give, through better constraints or observed excesses, a sharpening of where to look for this Higgs.  In that regard, it is important to assess the prospects of looking for such Higgses at the Tevatron and LHC.  Hadron colliders are decidedly more challenging experiments in which to look for Higgses, but the luminosity gains over LEP2 do allow searches to focus on rarer modes that LEP could not search for.  We will focus on the LHC capabilities, but searches at Tevatron could be interesting as well. 
%\disc{We don't say anything about Tevatron.} \disc{Hmmm, true.}

The first possibility is to focus on the suppressed Standard Model Higgs decay modes, which now require extra luminosity in order to discover the reduced branching ratio.  Maintaining $5\sigma$ discovery for these statistics limited searches requires a naive increase in integrated luminosity by a factor of $1/BR(h\to SM)_{new}^2$.  Since the LEP2 limit requires this branching ratio to be below $25\%$, the required luminosity is at least 16 times larger.  For instance extrapolating the latest CMS TDR analysis of $h \to 2\gamma$  \cite{cmstdr}, this pushes the necessary luminosity to $\gtrsim 250$ fb$^{-1}$.  This could be an overly optimistic extrapolation, as the background rates may increase as the LHC moves to design luminosity of $\sim 100 \; {\rm fb}^{-1}/{\rm year}$.  At any rate, this seems to require a few years of running at design luminosity and in more extreme suppressions of the SM branching ratio to an SLHC upgrade amount of luminosity \cite{Gianotti:2002xx}.  Instead, it is interesting to consider the prospects of detecting the new nonstandard decay modes themselves since their branching ratio is still $O(1)$.  For the purposes of the rest of this section, we will assume that the branching ratios are set by  the offshell W and Z decays.  Since the most promising modes are those containing leptons or neutrinos, in a particular model with larger (smaller) leptonic branching ratios will have a larger (smaller) reach at LHC.  

The new modes that are particularly interesting are the leptonic decays of both the offshell $W$ and $Z$'s.  In particular, the topologies with light charged leptons are particularly interesting.  In the charged decay (type $ii$), looking at the topology of a hadronic tau with a electron (or muon) decay of the $W^*$ is very similar to the standard $h\to 2\tau$ search.  This is at a higher branching ratio since the SM search requires using the Higgs $2\tau$ decay branching ratio which is about 8\%.  However, it is important to note that certain assumptions used in the $h\to 2\tau$ analysis do not apply.  For instance, since the taus from SM Higgs decays are highly boosted, the neutrino momenta can be assumed to point along the directions of the visible decay products.  This allows a Higgs mass reconstruction which would not apply for the nonstandard decay.  In the neutral decays (type $i$), the light charged lepton decay is small because of the 6\% branching ratio of the $Z^*$, but the overall branching ratio is still large, so this could still be a good channel.  Note that some older SUSY Higgs studies considered $H, A \to \chi_1 \chi_2 \to 2\chi_1 Z^{(*)}$ and found that the prospects were not promising \cite{Baer:1992kd}.  However, to our knowledge no study has considered weak boson production of such a Higgs cascade, since the mass spectrum of neutralinos generically prevents the Higgslike scalar in the MSSM from decaying into these neutralinos \footnote{For instance, GUT gaugino mass relations coupled with the chargino limit, force the lightest neutralino to be heavier than 50 GeV, which gives very little phase space for the nonstandard decay.}.  Hopefully, tagging on the forward jets in the weak boson fusion production will enhance signal over background to detect such a decay.  The Higgs production with an associated $W$ is also interesting as it generates a trilepton signal. The starting cross section is about 3 pb and with leptonic decay of the W gives a cross section of about 30 fb.  If the efficiency of this is reasonable, this trilepton signal could be seen over background.  These lepton signals are currently in investigation  \cite{spencertomas}.             

The fact that these leptonic modes resemble some classic supersymmetric signals introduces some interesting analysis issues.  If this Higgs appears in a supersymmetric theory, it will be necessary to produce cuts to isolate the Higgs component from the sparticle production component (for e.g., see \cite{Baer:1992kd}).  However, if this Higgs appears in a nonsupersymmetric theory, supersymmetric searches should pick it up as long as some non-optimized selection cuts are applied.  Ultimately this could require some interplay between experimentalists doing Higgs searches and those looking for supersymmetry, so any unexpected excesses inconsistent with the analyses original intent should be closely scrutinized.    

There also some interesting associated channels.  An associated invisible decay signal comes from when the $Z^*$ decays invisibly, which is expected to be of order 20\% which is the $95 \%$ CL sensitivity of LHC for 30 fb$^{-1}$ \cite{atlasinv,cmsinv}; so it could be feasible to detect this within the design luminosity running of LHC.  On the other hand, the decays with jets plus missing energy have large backgrounds from $W,Z$ production in association with jets.  It would take an indepth analysis to determine if the Higgs decay can be separated from this background.  Perhaps the onshell scalar could be an interesting possibility if the fact that the $b$ jets form a rough mass peak is useful.  
%The fact that $h\to 2b$ cannot be searched  for suggests that this isn't possible.  
There is missing energy in the decay, which effectively makes it look like a Higgs produced in association with a $Z$, so maybe there is some small hope.  So it remains to be seen if this channel could prove useful

Overall, there are many decay channels of interest, with potentially no one channel being sufficient.  In this scenario, the SM branching ratio is suppressed to the extent that a large amount of integrated luminosity is required for discovery.  However, there are searches in missing energy channels that could be modified to pick up the nonstandard decays.  The topologies of $l \bar{l} +\met; \tau \bar{l}+\met;{\rm trilepton} +\met; {\rm invisible}$ are all interesting possibilities to search for such a Higgs and could be picked up by altering some standard supersymmetric searches (squark cascades, trilepton searches, etc...).  This scenarios's correlation in signals could be helpful in boosting the significance of the search and hopefully maintain LHC's discovery potential for Higgs bosons.

\section{Conclusions \label{sec:conclusions}}
There are many reasons, both experimental and theoretical, to consider nonstandard Higgs decays as a means to allow a lighter Higgs.  This scenario  implies that the dominant Higgs decay is nonstandard, while Higgs production cross sections are unmodified.  In this note, we have considered in detail a new type of decay containing both visible and missing energy, in particular, decays of the form $h \rightarrow f \bar f + \me$ and $h \rightarrow \tau f \bar{f}' +\me$. Such decays can arise naturally in many theories beyond the Standard Model.  The phenomenology is simply described in terms of the language of On-shell Effective Theories, which is a natural way to focus on the physics without discussing model realizations.  In this way, we have been able to deduce what limits LEP searches put on such decays.  

In general, most decays (those not involving $b$ quarks or light charged leptons) are allowed with order one branching ratio for a 100 GeV Higgs, which has significance for precision electroweak studies, as well as for the possibility of less tuned supersymmetric theories.  Cases of $f \bar f + \me$ decays of the Higgs in particular appear allowed to $\sim 104\; \gev$ if the branching ratios for $f \bar f$ are those of an offshell Z, and could be  as low as $\sim 98\; \gev$\ if the purely invisible decays are suppressed, for instance by having the Higgs decay with 20\% SM modes and 80\% into $f \bar f \me$. One case which appears highly constrained is when $f \bar f = b \bar b$ exclusively, which can arise with when decays are mediated by an onshell scalar that itself decays primarily into bottom quarks.  In this case, a branching ratio of $\sim 25\%$ is the maximum for a 100 GeV Higgs.   Cases where the Higgs decays into a new sterile neutrino which decays via offshell $W^*,Z^*$ appear allowed for Higgses as light as $100-104\; \gev$.   This last scenario is realized in a simple example, via a two-Higgs doublet model where a weak scale sterile neutrino mixes with the Standard Model tau neutrino, with a mixing angle of $O(10\%)$.  Further studies are being performed on supersymmetric models and will be presented elsewhere \cite{UsandDavid}.  Potential searches for such nonstandard decays can be done at Tevatron and LHC, primarily in channels with leptons plus missing energy.  There could be interesting analysis issues if these are picked up in early stages of traditional supersymmetry searches (for e.g., trilepton searches), motivating further investigation into any excesses.  Overall, the discovery capabilities probably will rely on combining many of these nonstandard channels along with the usual Higgs search channels, but the feasibility remains uncertain and requires further study.

\vskip 0.15in

\section*{Acknowledgements} The authors thank Tilman Plehn, Liantao Wang for useful conversations.  NW thanks the Stanford and SLAC theory group for hospitality during the completion of this work.
The work of S. Chang and N. Weiner was supported by NSF CAREER grant PHY-0449818 and DOE grant \# DE-FG02-06ER41417. 
\bibliography{hx2x1}
\bibliographystyle{apsrev}

\end{document}